# Visualizing Object-oriented Software for Understanding and Documentation

Ra'Fat AL-msie'deen

Department of Information Technology, Mutah University, Al-Karak, Jordan

*Abstract*—**Understanding or comprehending source code is one of the core activities of software engineering. Understanding object-oriented source code is essential and required when a programmer maintains, migrates, reuses, documents or enhances source code. The source code that is not comprehended cannot be changed. The comprehension of object-oriented source code is a difficult problem solving process. In order to document object-oriented software system there are needs to understand its source code. To do so, it is necessary to mine source code dependencies in addition to quantitative information in source code such as the number of classes. This paper proposes an automatic approach, which aims to document object-oriented software by visualizing its source code. The design of the object-oriented source code and its main characteristics are represented in the visualization. Package content, class information, relationships between classes, dependencies between methods and software metrics is displayed. The extracted views are very helpful to understand and document the object-oriented software. The novelty of this approach is the exploiting of code dependencies and quantitative information in source code to document object-oriented software efficiently by means of a set of graphs. To validate the approach, it has been applied to several case studies. The results of this evaluation showed that most of the object-oriented software systems have been documented correctly.**

## I. INTRODUCTION

Different studies about the software understanding indicate that programmers rely on good software documentation [22]. Software rapidly becomes very complex when its size increases, which make its development very hard task. The very huge amount of information represented in the software source code, at all granularity levels (*i.e.,* package, class, attribute and method) make understanding and documenting software a very difficult, lengthy, and error-prone task [3]. Moreover, manually-written documentation is not feasible for being incomplete, either because it is very time-consuming to create, or because it must frequently be updated [12]. This paper proposes a new approach called Vsound[1] to understand and document the object-oriented (OO) software by visualizing its source code as a set of graphs at all granularity levels. In order to give a precise definition of the software documentation, Vsound considers that the software documentation is the process of taking software source code and understanding it by visualizing its source code as a set of graphs.

Software visualization is the use of the crafts of typography, graphic design, animation, and cinematography with modern human-computer interaction and computer graphics technology to facilitate both the human understanding and effective use of computer software [13]. Software visualization (*resp.* software documentation) can tackle three different types of aspects of software (*i.e.,* static, dynamic and evolution) [2]. The visualization of the static aspects of software focuses on visualizing software as it is coded. While, the visualization of the dynamic aspects of software represents information about a specific run of the software and helps comprehend program behavior and, at last, the visualization of the evolution of the static aspects of software adds the time factor to the visualization of the static aspect of software. This paper tackles only the visualization (*resp.* documentation) of the static aspects of software.

Software comprehension[2] is the process whereby a software practitioner understands a software artifact using both knowledge of the domain and/or semantic and syntactic knowledge, to build a mental model of its relation to the situation [14]. Software understanding is one of the main software engineering activities. Software understanding is the process of taking software source code and understanding it. Software comprehension is necessary when a programmer migrates, reuses, maintains, documents or enhances software systems. Software that is not comprehended cannot be changed [1]. The domains of software documentation and visualization are driven by the need for program comprehension. Software visualization (*resp.* documentation) is a successful software comprehension way. Software comprehension is an important part of software evolution and software maintenance.

The software maintenance process is the most expensive part of software development. Most of time spent in software maintenance is used to comprehend the software code and the instructions that have to be changed [17]. Software maintenance is the modification of a software product after delivery to correct faults, to improve performance or other attributes, or to adapt the product to a changed environment [16]. The software undergoes modification to source code and related documentation due to a problem or the need for enhancement. The goal is to modify the existing software while preserving its integrity [15].

---

[1] Vsound stands for <u>V</u>isualizing object-oriented <u>SO</u>ftware for <u>UN</u>derstanding and <u>D</u>ocumentation.

[2] *a.k.a.,* "program understanding" or "source code comprehension".

Software must frequently evolve to adjust to new features (*resp.* environments) or to meet specific requirements. Software system wants to evolve in order to be used longer time. The company often evolves frequent release of new versions of the original software. Each release results in the increase of software system size and complexity. Thus, software implementation that facilitates modify is key for reducing maintenance costs and effort. Software evolution reflects the process of progressive change in the attributes of the evolving entity or that of one or more of its constituent elements [20]. In other words, software evolution is linked to how software systems evolve over time.

Software reuse is important to reduce the cost and time of software development. In order to reuse existing source code there is a need to understand and document the software code. With increasing of software complexity (*i.e.,* increasing count of lines of code) the need for reuse grows. Software reuse is the process of creating software systems from existing software rather than building software systems from scratch [18]. Software reuse helps reduce the development and maintenance effort. In addition, software reuse improves software quality and decrease time-to-market [38].

Software comprehending is very necessary for software changes in the maintenance stage. The changes to the software's code may be affected to another part of the code. These situations make the developer spent more time and effort to find the affected lines of the whole code. Change impact analysis is the process of identifying the potential consequences of a change, or estimate what needs to be modified to accomplish a change [21]. Change impact analysis support program understanding by finding the potential effect or dependency information in source code.

This paper proposes a new approach, which aims to document software systems by visualizing their code. The documentation process is very useful for software understanding, maintenance, evolution, reuse and changes. Documentation process involves the creation of alternative representations of the software, usually at a higher level of abstraction. It also involves analyzing the software in order to determine its elements and the relations between those elements. Software visualization is commonly used in the fields of reverse engineering and maintenance, where huge amount of code need to be understood [23]. Reverse engineering is the process of analyzing a subject system to identify the system's components and their interrelationships and create representations of the system in another form or at a higher level of abstraction [16].

To assist a human expert to document software system, this paper proposes an automatic approach, which generates a set of graphs (*i.e.,* documents) using source code elements of a software system. Compared with existing work that documents source code (*cf.* section *related work*), the novelty of Vsound approach is that it exploits source code dependencies and its quantitative information, to document OO software in an efficient way by means of a set of graphs. Vsound accepts as input the source code of software as a first step. Then, based on the static code analysis, Vsound generates an XML file contains the main source code elements (*e.g.,* package, class attribute and method) and the dependencies (*e.g.,* inheritance) between those elements. Next, Vsound documents the software by extracting a set of graphs based on the source code; each graph considers as a document. The mined documents cover all granularity levels (*i.e.,* package, class, attribute and method) of the source code.

The Vsound approach is detailed in the remainder of this paper as follows. Section 2 briefly presents the background needed to understand the proposal. Section 3 shows an overview of Vsound approach. Section 4 presents the software documentation process step by step. Section 5 describes the experiments that were conducted to validate Vsound proposal. Section 6 discusses the related work, while section 7 concludes and provides perspectives for this work.

## II. BACKGROUND

This section provides a glimpse on software documentation and visualization. It also shortly describes dependencies between source code elements which consider relevant to Vsound approach.

The software documentation process aims to generate documents with abstract information based on software source code. The extracted documents are very useful, especially when software documents are missing. The good software documentation process helps the programmer working on software to understand its features and functions. The software documentation process is specific, where it provides all the information important to the person who works on the software at different levels of abstraction. The documentation process aims to translate source code of the software system into a set of documents (*i.e.,* graphs). In reality, several software systems have little to no software documentation, especially the legacy software. Many companies are facing some problems with legacy systems such as: software understanding and software maintenance. The reason behind these problems is the absence of software documentation. Software systems that are not documented hard to be changed [41]. Common examples of such documentation include requirement and specification documents [22]. Vsound provides as output a set of documents describe the source code and its dependencies.

Software visualization is the graphical show of information about the software source code. Software visualization is not a simple process since the amount of information to be included in the graph is may be far bigger than can be displayed. The software visualization tool presents information about the software source code at different levels of abstraction. The visualization tool focuses on displaying different aspects of the source code. It should provide a way to choose and display just particular information based on the software source code. Usually, the visualization process must find out the level of abstraction of the information it presents about the software source code. The software visualization tool must convert the software source code into a graph. It also must able to visualize a huge amount of information regarding the software source code. Moreover, the visualization tool must provide an easy way for navigation [24].

Vsound relies on the source code of software systems to extract a set of documents describing the software. The software source code is the most important resource of information when the software documentation is missing. In order to document existing OO software based on its source code there is a need to extract the main source code elements and dependencies between those elements. Vsound considers that the software implementation consists of OO source code elements and the code dependencies. The main elements of OO source code include: package, class, attribute and method. Vsound also focuses on the method body. The method body consists of a method signature, which represents the access level, returned data type and parameter list (*i.e.,* name, data type and order). It also consists of the body of the method (*i.e.,* local variable, method invocation and attribute access). Vsound focuses on the main dependencies between source code elements such as: inheritance, attribute access and method invocation in the documentation process. Inheritance relation occurs when a general class (*i.e.,* superclass) is connected to its specialized classes (*i.e.,* subclasses). Method invocation relation occurs when methods of one class use methods of another class. While, attribute access relation occurs when methods of one class use attributes of another class.

## III. APPROACH OVERVIEW

This section provides the main concepts and hypotheses used in the Vsound approach. It also gives an overview of the software documentation process. Then, it presents the OO source code model. Finally, it shortly describes the example that illustrates the remaining of the paper.

### A. Key Ideas

The general objective of Vsound approach is to document the source code of a single software system based on the static analysis of its source code. Mining the main entities of the source code in addition to the source code dependencies is a first necessary step towards this objective. Vsound considers software systems in which software functionalities are implemented at the programming language level (*i.e.,* source code). Vsound also restricts to OO software system. Thus, software functionalities are implemented using OO source code elements such as packages, classes, attributes, methods or method body elements (*i.e.,* local variable, attribute access, method invocation). There are several ways to document the software source code such as generate a descriptive comments that summarize all software classes or methods [12]. Vsound aims to document the software source code as a set of documents (*i.e.,* graphs). The software documentation process via single graph is difficult. Thus, there is a need to document the software source code through several documents with details.

The documentation process must cover all granularity levels of the software source code (*i.e.,* package, class, attribute and method). Vsound focuses on the extracting of three types of documents (*i.e.,* graphs). The first document contains general information about the software source code. This document called the package document. It represents all software packages in addition to the number of classes per package. It also provides quantitative information about a software system. The second document contains information about software classes. This document called the class document. It represents all information about the class, such as the number of attributes and methods in addition to class dependencies. The third document contains information about software methods. This document called the method document. It represents all information about the method, such as the number of parameters and local variables in addition to method dependencies like attribute access and method invocation.

### B. The OO Software Documentation Process

Vsound goal is to document the OO software by using the source code of this software. The software documentation process takes the software's source code as its inputs and generates a set of documents as its outputs.

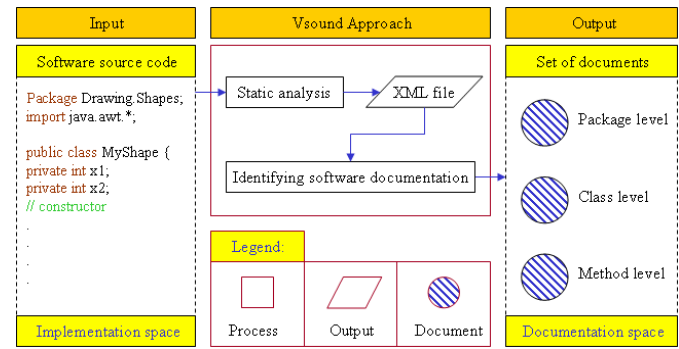

Figure 1. The OO software documentation process.

Vsound approach exploits the main source code elements and the dependencies between those elements in order to document and understand existing software system. Figure 1 shows the software documentation process. The first step of this process aims at extracting the main OO source code elements (*i.e.,* package, class, attribute, method) and their relationships (*i.e.,* inheritance, method invocation and attribute access) based on the static analysis of source code. In the second step, Vsound approach relies on the mined source code to document software at package level. In the third step, Vsound approach documents the software at class level based on the extracted source code. The last step of this process aims at documenting the software at method level. Finally, these documents (*i.e.,* graphs) are used to understand and document the software system.

### C. Object-oriented Source Code Model

The Vsound source code meta-model was inspired by the FAMIX [39] information exchange meta-model. The source code meta-model (*cf.* Figure 2) displays the main source code elements and their relations. Mainly, the reader gets enough information if he considers the main type entities that construct an object-oriented system. These are: package, class, interface, attribute, method, and the relations between them, such as inheritance, access and invocation. The OO source code model shows structural source code entities such as: packages, classes and methods. In addition, this model represents explicitly information such as: a class inherits from another class (*i.e.,* inheritance), a method accesses attributes (*i.e.,* access) and a method invokes other methods (*i.e.,* invocation). These abstractions are very important and needed for reengineering tasks such as: dependency analysis [40].

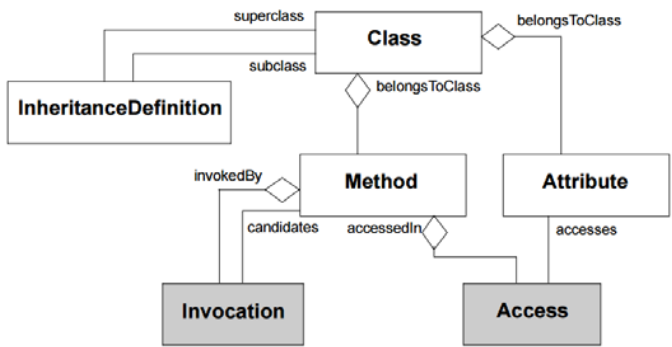

Figure 2.   The source code meta-model [39].

### D.   An Illustrative Example

As an illustrative example, this paper considers the drawing shapes software[3] (*cf.* Figure 3). This software allows a user to draw three different kinds of shapes. The drawing shapes software allows user to draw lines, rectangles and ovals and choose the color of the drawn shape (*cf.* Figure 3). This example used to better explain some parts of this paper. However, Vsound approach only uses the source code of software as input of the documentation process and thus do not know the code dependencies or software metrics in advance.

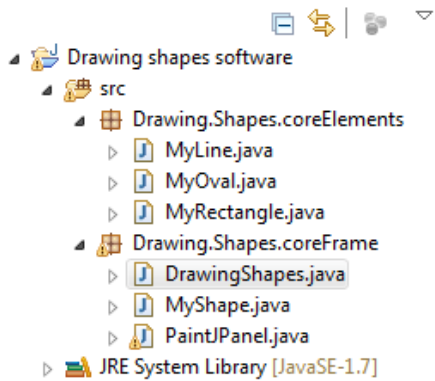

Figure 3.   The drawing shapes software.

## IV.   THE OBJECT-ORIENTED SOFTWARE DOCUMENTATION PROCESS STEP BY STEP

This section describes the OO software documentation process step by step. According to Vsound, the approach identifies the software documentation in four steps as detailed in the following.

### A.   Extracting the Source Code of Software via Static Code Analysis

Static code analysis is the analysis of computer software that is performed without actually executing programs built from that software [4]. While, analyzing the actions performed by a program while it is being executed is called dynamic analysis [5]. Vsound approach accepts as input the source code of software. Then, the proposed approach generates an XML file based on the static code analysis[4]. The mined XML file contains structural information between OO source code elements (*e.g.,* draw method is inside the MyLine class). It also contains structural dependencies between source code elements (*e.g.,* inheritance, attribute access and method invocation). The Eclipse Java Development Tools (JDT) and the Eclipse Abstract Syntax Tree (AST) can be used to access, modify and read the elements of a Java program [37]. ASTs are broadly used in several fields of software engineering. AST is used as a representation of source code [25]. The extracted XML file contains all information needed to document the software by visualizing its code as a set of graphs. Figure 4 shows the mined XML file from the source code of drawing shapes software.

```
<Project ProjectName="Drawing shapes software" LinesOfCode="1500">
  <Packages>
    <Package PackageName="Drawing">...</Package>
    <Package PackageName="Drawing.Shapes">...</Package>
    <Package PackageName="Drawing.Shapes.coreElements">
      <Classes>
        <Class ClassName="MyLine" classAccessLevel="public">
          <SuperInterfaces/>
          <Attributes/>
          <Methods>
            <Method MethodName="MyLine" MethodAccessLevel="public">
              <Parameters NumberOfParameters="5">...</Parameters>
              <LocalVariables/>
              <AttributeAccesses>...</AttributeAccesses>
              <MethodInvocations/>
              <MethodExceptions/>
            </Method>
            <Method MethodName="draw" MethodAccessLevel="public">
              <Parameters NumberOfParameters="1">...</Parameters>
              <LocalVariables/>
              <AttributeAccesses>...</AttributeAccesses>
              <MethodInvocations>...</MethodInvocations>
              <MethodExceptions/>
            </Method>
          </Methods>
        </Class>
        <Class ClassName="MyOval">...</Class>
        <Class ClassName="MyRectangle">...</Class>
      </Classes>
    </Package>
    <Package PackageName="Drawing.Shapes.coreFrame">...</Package>
  </Packages>
</Project>
```

Figure 4.   The extracted XML file of drawing shapes software.

### B.   Identifying the Package Document

Vsound approach extracts several documents based on the software source code. These documents cover all granularity levels of the source code. To provide a better understanding of existing software, it is impossible to gather all information in one graph. Vsound approach provides one document (*i.e.,* graph) for every granularity level (*i.e.,* package, class, attribute, method). The package document aims to provide specific information about software. The package document contains

---

information about software packages and quantitative information about the software (*cf.* Figure 5).

Figure 5 shows the mined package document from drawing shapes software. This document provides information about software packages. As an example, from the graph in Figure 5, drawing shapes software consists of two packages, each one contains three classes. In addition, this document provides information about software metrics such as Lines of Code (LoC), Number of Packages (NoP), Number of Classes (NoC), Number of Attributes (NoA) and Number of Methods (NoM). As an example, from the graph in Figure 5, drawing shapes software consists of 29 methods and 14 attributes.

In order to understand the software source code, the package document is very useful. The goal of this document is to give a general view about software system and present the size of the software (*i.e.,* large, medium or small system). Vsound approach applies to different sizes of software systems. The different complexity levels show the scalability of Vsound approach to dealing with such systems.

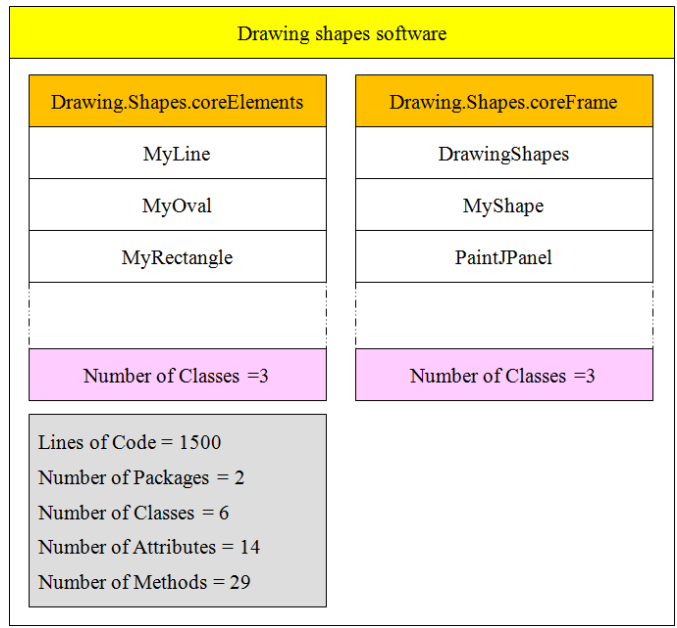

Figure 5.   The package document mined from drawing shapes software.

## C.   Identifying the Class Document

The class document displays information about software classes. This information is very helpful toward understanding the software. Vsound approach identifies three documents belong to the class document category. The first document represents the class information document (*cf.* Figure 6). This document shows the number of classes per package. It also presents information about each class in the package. The class information document consists of the class name, super class name, is an interface, super interface name, number of attributes and number of methods. As an example, the class MyShape in Figure 6 consists of five attributes and 12 methods.

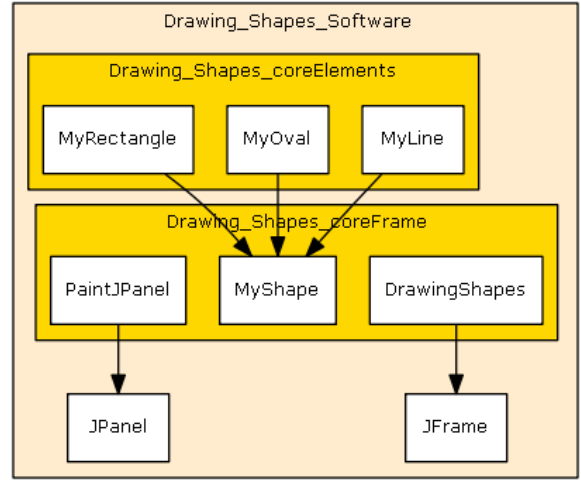

Figure 6.   Part of the class information document mined from drawing shapes software.

The second document represents the class dependency document (*cf.* Figure 7). This document shows the main relations between software classes (*i.e.,* Inheritance relation). As an example, the classes MyLine, MyOval and MyRectangle in Figure 7 have a super class called MyShape.

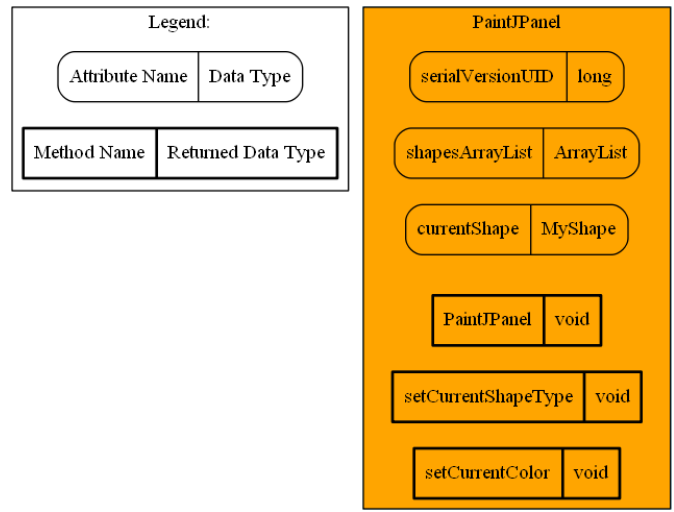

Figure 7.   The class dependency document mined from drawing shapes software.

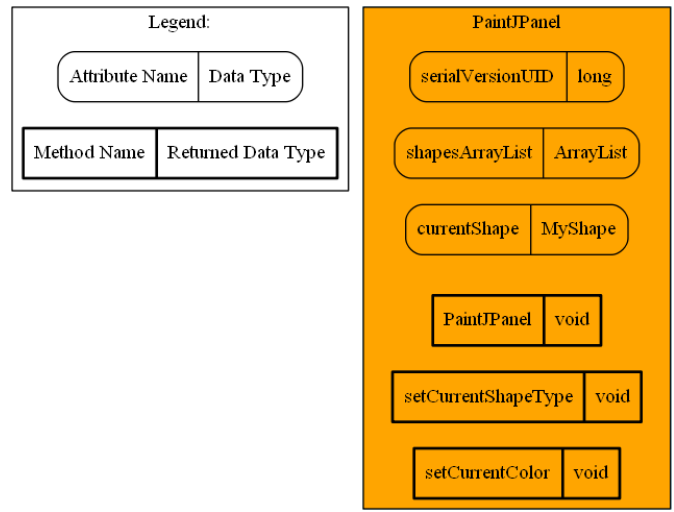

Figure 8.   Part of the class content document mined from drawing shapes software.

The third document represents the class content document (cf. Figure 8). This document shows the main content of each class. It also shows the size of class, where the height of class considered as an indicator of class size. Vsound considers the main elements in the classes. The class content document includes the attribute name and its data type in addition to the method name and its returned data type. As an example, the class PaintJPanel contains currentShape attribute of MyShape type. It also contains setCurrentShapeType method.

### D. Identifying the Method Document

The method document shows information about software methods. This information is very useful toward understanding the software system. Vsound approach identifies three documents belong to the method document category. The first document represents the method information document (*cf.* Figure 9). This document provides information about software methods. The method information document contains the following information: the name of a method, the method returned data type, is a static method, number of parameters and the parameter list (*i.e.*, name, data type and order). As an example, the method MyRectangle in Figure 9 consists of 5 parameters.

The second document represents the method content document (*cf.* Figure 10). This document shows the main content of each method. It also shows the size of method, where the height of method can considered as an indicator of method size. Vsound considers the main elements in the method body. The method content document includes the local variable name and its data type, attribute access name and its type and method invocation name and the declared class. As an example, method PaintJPanel in Figure 10 contains addMouseListener which is a method invocation.

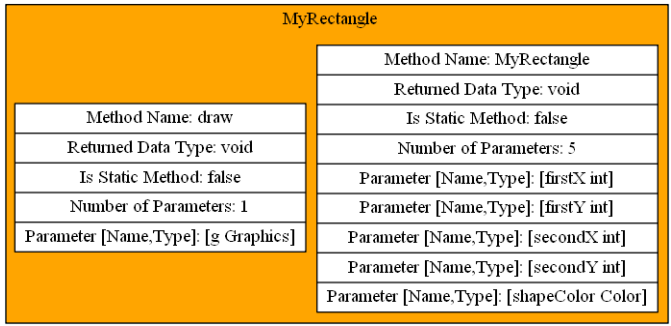

Figure 9.  Part of the method information document mined from drawing shapes software.

The third document represents the method dependency document (*cf.* Figure 11). This document shows the main relations between software methods (*i.e.,* method invocation and attribute access). As an example, the method "draw" that is declared in MyShape class is invoked by the paintComponent method in PaintJPanel class. Also, the currentShape attribute that declared at DrawingShapes class is accessed in paintJPanelMouseDragged method that declared at PaintJPanel class.

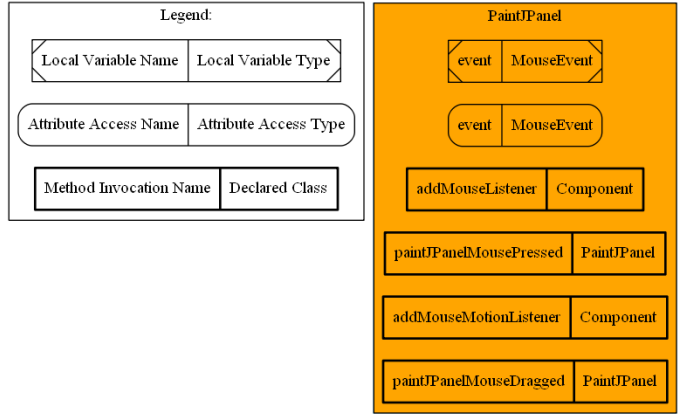

Figure 10.  Part of the method content document mined from drawing shapes software.

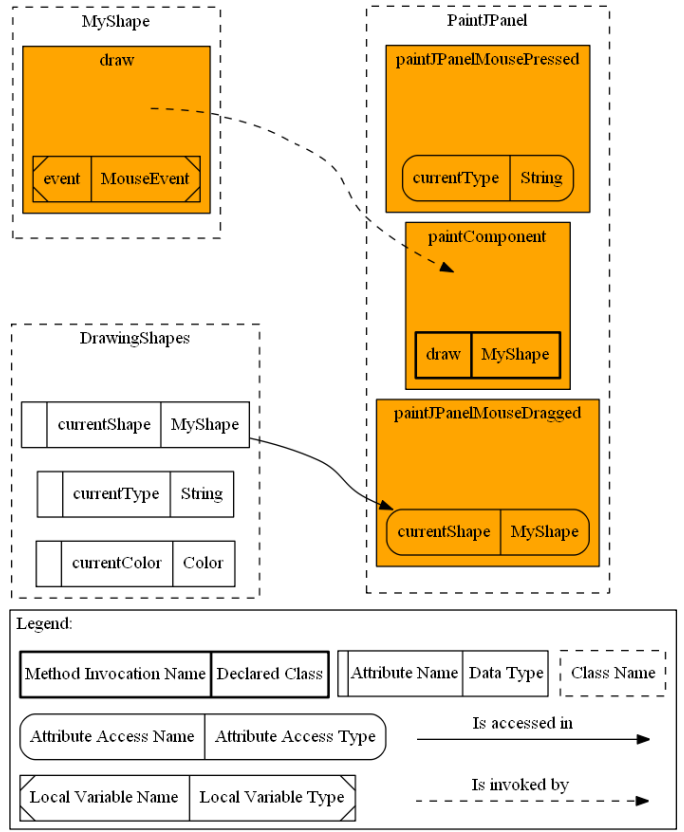

Figure 11. Part of the method dependency document mined from drawing shapes software.

## V.  EXPERIMENTATION

This section presents the experiment that conducted to validate the Vsound approach. Firstly, this section presents the ArgoUML case study. Next, it presents the evaluation metrics. Then, it also presents detail the architecture and functioning of the Vsound prototype tool and, at last, it presents the software documentation results and threats to validity of Vsound approach.

## A. Case Study

In addition to the toy drawing shapes example used in this paper, Vsound approach has been tested on other software called ArgoUML. ArgoUML is a widely used open source tool for UML modeling tool. ArgoUML supports the following UML 1.4 diagram types: class diagram, state chart diagram, activity diagram, use case diagram, collaboration diagram, deployment diagram and sequence diagram. The advantage of using the ArgoUML as a case study is that ArgoUML software well documented. In this evaluation, the results are based on the source code of the software that is freely available for downloading in the case study website[5]. ArgoUML runs on any Java platform and is available in ten languages. ArgoUML software is presented in Table 1 characterized by metrics LoC, NoP, NoC, NoA and NoM.

TABLE I. SIZE METRICS FOR ARGOUML SOFTWARE SYSTEM.

| Software | LoC | NoP | NoC | NoA | NoM |
|----------|-----|-----|-----|-----|-----|
| ArgoUML | 120,348 | 81 | 1,666 | 3977 | 14904 |

## B. Evaluation Measures

In order to evaluate Vsound approach, precision and recall measures are used. In this paper, the precision is the percentage of correctly retrieved links to the total number of retrieved links. The recall is the percentage of correctly retrieved links to the total number of relevant links [36]. In this work, link means: source code element (package, class, attribute, method or local variable) or dependency (inheritance, method invocation or attribute access). All measures (*i.e.*, precision and recall) have values in [0, 1]. If precision is equal to one, every one of retrieved links is relevant. However, relevant links might not be retrieved. If recall is equal to one, all relevant links are retrieved. Nevertheless, some retrieved links might not be relevant. For example, by considering software system contains 95 relations (*i.e.,* inheritance relation). After applying the proposed approach on this software, the result shows that 90 relations are identified correctly (*i.e.,* all retrieved links are relevant). However, five relations are missing. In this case, precision is equal to 1 (*i.e.,* 90/90=1) and recall is equal to 0.94 (*i.e.,* 90/95=0.94).

## C. A Simplified Structural View of the Architecture of Vsound

The developed prototype tool [6] of Vsound approach implements the proposed software documentation process. Figure 10 provides an overview of the Vsound tool architecture. It receives as input the software source code and produces as output a set of documents. The XML_Generator module of Vsound produces an XML document which represents the software source code elements and the dependencies between those elements. The tool then starts to generate the software documents by using the DOT_File_Builder component. These documents are serialized as DOT files. DOT is a plain text graph description language. Starting from these DOT files, the tool builds the software document (*i.e.,* graph). For doing so, the Vsound_GUI component of Vsound uses an external library, called Graphviz[7] to produce SVG files containing the documentation of each software artefact (*i.e.,* package, class and method). SVG (Scalable Vector Graphics) is an XML-based file format for describing vector graphics [28].

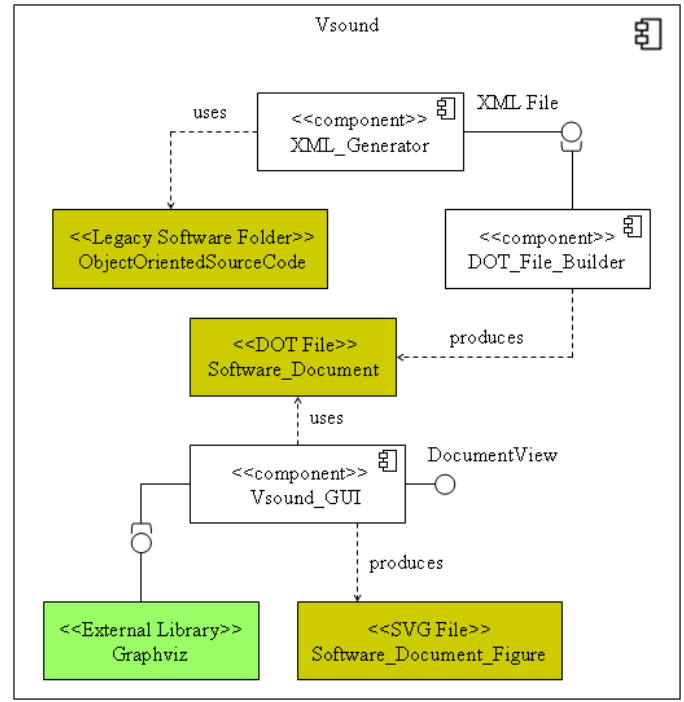

Figure 12. A simplified structural view of the architecture of Vsound.

## D. Result

Vsound approach has tested on the ArgoUML case study and obtained promising results. The preliminary evaluation of Vsound shows the significance of this approach. Vsound approach extracted a collection of documents from the source code of ArgoUML software. The first document represents the package document which represents general information about ArgoUML such as: number of packages, number of classes per each package and quantitative information about ArgoUML like LoC and NoM. The second document represents the class document (i.e., the class information document, the class dependency document and the class content document). The identified documents by Vsound provide useful information about software classes. The third document represents the method document (i.e., the method information document, the method content document and the method dependency document). This document shows meaningful information about the software methods. Results show that precision appears to be high for all mined documents from ArgoUML software. This means that all mined documents grouped as software documentation are relevant. Considering the recall metric, recall is also quite high. This means that most source code elements and their dependencies that compose software documentation are mined. Thanks to Vsound approach that identifies software documentation in a novel way.

---

*E. Threats to Validity*

Vsound approach considers only the Java software systems. This represents a threat to prototype validity that limits Vsound implementation ability to deal only with software systems that are developed based on the Java language. Vsound assumes that source code elements and the dependencies between those elements can be determined statically, such as ArgoUML used in Vsound evaluation. Nevertheless, there exist systems that only behave differently depending on runtime parameters. For such systems, there is a need to extend Vsound to include dynamic analysis techniques. Vsound approach assumes that software documentation can be determined graphically as a set of documents based on the source code. In some cases, there is a need to document software systems by describing their functionalities. This means that Vsound maybe not reliable (or should be improved with other techniques) in all cases to identify software documentation. Documenting software using the names of source code elements and dependencies in its implementation is not always reliable. Thus, it is important to use the comments (*i.e.,* line and block comment) in the source code in order to enhance the documentation process.

## VI. RELATED WORK

This section presents the related work relevant to Vsound contribution. It also provides a concise overview of the different approaches and shows the need to propose Vsound approach.

Wettel *et al.,* [11] proposed an approach to visualize object oriented software as a city to solve the navigation problem. CodeCity is a visualization tool that represents the software with a city metaphor. The classes are represented as buildings and the packages as districts. Moreover, some of the visual properties (*i.e.,* width, height, position, and color) of the city artifacts carry information about the software element they represent (*e.g.,* the height of the building represents the number of methods). Their approach does not consider dependencies between source code elements.

Hammad *et al.,* [32] used software visualization techniques to visualize class coupling based on analyzing the source code statically. Other works such as [33] and [34] applied software visualization to model the dynamic behavior of the software by instrumenting the source code in order to monitor the program executions. Moreover, visualization techniques can be applied on software documentation to make it easier and more helpful. For example, work in [35] proposed a visualization framework to visualize bug reports that are saved in software repositories. The proposed framework visualized the status changing for selected bugs, as well as, bug-developer relations by using different shapes and colors.

Al-msie'deen *et al.,* [6] [9] [29] [30] proposed an approach called REVPLINE to identify and document features from the object-oriented source code of a collection of software product variants. The authors presented a new way to document the mined feature implementations in [19]. The proposed approach gives as output for each feature implementation, a name and description based on the use-case name and description [7] [8]. REVPLINE approach aims to document the extracted features from a collection of software variants, while Vsound aims to document OO software as a set of graphs. Hammad *et al.,* [31] presented an approach that focuses on analyzing code changes to automatically detect any OO constraints violation without using graphs as in Vsound approach.

Graham *et al.,* [26] used a solar system metaphor to represent the software source code. The main code entities (*i.e.,* packages and classes) are represented as planets encoding software metrics in the planets' size and color. This visualization represents the software as a virtual galaxy consists of many solar systems. Each solar system represents a package. The central star of each solar system represents the package itself, while the planets, in orbit around it, represent classes within the package. For planets, the blue planets represent classes while the light blue planets represent interfaces. Solar systems are shown as a circular formation. The Solar System metaphor represents packages and classes of software without considering methods and their body. In addition, it focuses only on the inheritance relation between classes without considering the attribute access and method invocation relations.

McBurney and McMillan [12] create summaries of Java methods by using local information (*i.e.,* keywords in the method) and contextual information (*i.e.,* keywords in the most important referenced methods). The method's summaries are generated from elements in the method's signature and body. The approach applied only to the method without considering other artifacts (*e.g.,* package and class). The proposed approach used natural language processing and information retrieval techniques. The type of summary classified under abstract Summary.

Haiduc *et al.,* [27] proposed an approach for summarizing the software source code. The source code summaries consider only the software methods and classes. The proposed approach used information retrieval techniques such as latent semantic indexing and vector space model. The approach extracts the text from the source code of the software and converts it into a corpus (*i.e.,* source code corpus creation). Then, the approach determines the most relevant terms for documents in the corpus and includes them in the summary (*i.e.,* generating source code summaries using text retrieval). Their approach does not exploit structural information from the source code.

## VII. CONCLUSIONS AND FUTURE WORK

This paper proposed a new approach for documenting and understanding the software system. The novelty of this approach is that it exploits code dependencies and quantitative information in source code to document OO software in an efficient way by means of a set of documents. The Vsound approach has implemented and evaluated its results on several case studies. The results of this evaluation showed that most of the software systems were documented correctly. Regarding future work, Vsound approach plans to automatically generate descriptive comments that summarize all software packages, classes and methods. It also plans to use the line and block comments in the documentation process.

AUTHORS PROFILE

**Ra'Fat Al-Msie'Deen** is a lecturer at the Information Technology department in Mu'tah University, Al Karak - Jordan. He received his PhD. in Computer Science from University of Montpellier 2, France in 2014. He received his Masters degree in Information Technology from University Utara Malaysia, Malaysia in 2009. He got his B.Sc. in Computer Science from Al - Hussein Bin Talal University, Jordan in 2007. His research interest is software engineering with focusing on software documentation, software visualization, software maintenance, software evolution, reverse engineering, reuse, re-engineering, software product line and formal concept analysis.